\documentclass[letterpaper]{article}
\usepackage{aaai2026}
\nocopyright
\usepackage{times}
\usepackage{helvet}
\usepackage{courier}
\usepackage[hyphens]{url}
\usepackage{graphicx}
\usepackage{natbib}
\usepackage{caption}
\usepackage{subcaption}
\usepackage{booktabs}
\usepackage{amssymb}
\usepackage{mathtools}
\usepackage{amsthm}
\usepackage{algorithm}
\usepackage{algpseudocode}
\usepackage[capitalize,noabbrev]{cleveref}

\usepackage{amsmath,amsfonts,bm}

\def\eqref#1{equation~\ref{#1}}

\def\1{\bm{1}}

\DeclareMathAlphabet{\mathsfit}{\encodingdefault}{\sfdefault}{m}{sl}
\SetMathAlphabet{\mathsfit}{bold}{\encodingdefault}{\sfdefault}{bx}{n}

\algrenewcommand\algorithmicrequire{\textbf{Input:}}
\algrenewcommand\algorithmicensure{\textbf{Output:}}

\theoremstyle{plain}
\newtheorem{theorem}{Theorem}[section]
\newtheorem{proposition}[theorem]{Proposition}

\theoremstyle{definition}

\newtheorem{assumption}[theorem]{Assumption}
\theoremstyle{remark}

\title{OCO-S$^2$: Online Convex Optimization with Stateful Costs and Sparse Communication}

\author{
Yiwei Liu\textsuperscript{\rm 1},
Luwei Yang\textsuperscript{\rm 2},
Shunbo Lei\textsuperscript{\rm 1}
}
\affiliations{
\textsuperscript{\rm 1}School of Science and Engineering,
The Chinese University of Hong Kong, Shenzhen,
Shenzhen, Guangdong 518172, China\\
\textsuperscript{\rm 2}Center for Network and Machine Intelligent Optimization, Shenzhen Research Institute of Big Data,
Shenzhen 518172, China\\
yiweiliuswu@163.com,
yangluwei@sribd.cn,
leishunbo@cuhk.edu.cn
}

\begin{document}

\maketitle

\begin{abstract}
We study \textsc{OCO-S$^2$}, an online convex optimization setting in which
decisions drive a stable dynamical state, losses are incurred along the induced
state trajectory, and first-order feedback is available only through sparse
block communication with partial participation. This coupling creates a
dynamic-regret problem beyond pointwise OCO: the learner updates and holds
decisions at the block scale, whereas the hindsight comparator may vary at the
per-round scale. We propose \textsc{OCO-S$^2$-OGD}, a projected block online
gradient method that updates deployed decisions using sparse block-level
distributed feedback. We prove dynamic-regret bounds for the incurred
trajectory cost, quantifying the tradeoff among block communication,
comparator variation, state-memory truncation, and partial participation. We
further introduce a prediction-augmented variant,
\textsc{OCO-S$^2$-OGD-P}, and show that accurate block-level predictions improve
the optimization term in the regret bound through their realized
gradient-mismatch error. Overall, this work provides a regret-theoretic
foundation for communication-efficient online decision-making in systems where
algorithmic updates and physical state trajectories are intrinsically coupled.
\end{abstract}
\section{Introduction}
\label{sec:intro}

Online convex optimization (OCO) provides a standard framework for sequential
decision-making in nonstationary environments. At each round, a learner chooses
a feasible decision, observes loss information, and is evaluated by regret
against a hindsight benchmark. Dynamic-regret analysis further compares the
learner with path-length-bounded time-varying comparators, making it suitable
for environments with drifting optima
\cite{zinkevich2003ocp,hall2013trackingregret,jadbabaie2015dynamiccomparators,zhao2020dynamicconvexsmooth}.

In many online decision systems, however, the setting is not a pointwise OCO
problem with full per-round feedback. A deployed decision may drive an
operational state through system dynamics, so its effect persists beyond the
current round and later losses are incurred along the resulting state trajectory
\cite{li2021onlineaffineconstraints}. At the same time, distributed
implementations often cannot synchronize all clients at every round: feedback is
available only at block boundaries and may involve only a subset of clients
\cite{karimireddy2020scaffold}. The learner must therefore hold deployed
decisions within communication blocks, while the operational state continues to
evolve at the per-round time scale. This combination creates a trajectory-level
dynamic-regret problem: sparse communication affects not only the information
used for future updates, but also the state trajectory on which future costs are
incurred.

We study this coupled setting through \textsc{OCO-S$^2$}: online convex
optimization with stateful costs and sparse communication. In
\textsc{OCO-S$^2$}, sparse communication does not merely delay or restrict the
feedback available to the learner. Because deployed decisions are held within
communication blocks while the system state evolves at the per-round scale, the
communication protocol also changes the realized state trajectory itself and
hence the incurred cumulative cost.

This coupling calls for an analysis that separates algorithmic update error
from the trajectory error caused by block-held decisions. In particular, the
learner is evaluated on the state trajectory generated by its sparse
communication protocol, whereas the dynamic comparator is selected in hindsight
from a path-length-bounded class and may vary at the per-round scale. The regret
analysis must therefore account for the mismatch between a block-held online
trajectory and a per-round dynamic benchmark, in addition to the approximation
error caused by state memory and the sampling error caused by partial
participation.

To address these issues, we develop \textsc{OCO-S$^2$-OGD}, a projected block
online gradient method for \textsc{OCO-S$^2$}. The algorithm exploits the
fading-memory property of stable dynamics to build finite-memory surrogate
losses. After each communication block, only activated clients return local
surrogate gradients, and the server performs one projected update for the next
block. Thus the method communicates only \(O(T/K)\) times over a horizon of
length \(T\), while its analysis controls dynamic regret for the incurred
state-dependent cost.

The contributions of this paper are summarized as follows.
\begin{enumerate}
    \item We formulate \textsc{OCO-S$^2$}, an online convex optimization
    problem with stateful incurred costs and sparse communication. In this
    setting, decisions drive a stable dynamical state, while first-order
    feedback is available only through block-level partial communication.

\item We develop \textsc{OCO-S$^2$-OGD}, a projected block online gradient
method with sparse distributed feedback, and prove its dynamic-regret bound for
the incurred trajectory cost against path-length-bounded comparators. The bound
decomposes the effects of online descent, sparse communication, comparator
drift, state-memory truncation, and partial participation, while the method uses
only \(O(T/K)\) synchronization rounds over a horizon of length \(T\).

    \item We extend the framework to a prediction-augmented variant,
    \textsc{OCO-S$^2$-OGD-P}, and show that accurate block-level predictions can
    improve the optimization term in the regret bound.
\end{enumerate}
\section{Related Work}

\paragraph{Online convex optimization with memory.}
Online convex optimization with memory extends the classical OCO framework by
allowing the loss at each round to depend on a finite window of past decisions
\cite{zinkevich2003ocp,anava2013timeseries}. This formulation provides a
natural model for temporal coupling caused by switching costs, reconfiguration
costs, and finite-memory system dependence. Recent work has further incorporated
constraints and predictions into memory-based online convex optimization, where
losses and constraints may both depend on a finite decision window
\cite{abdullah2026coco}. This line is closest to ours because it moves beyond
pointwise losses and explicitly accounts for the effect of past decisions on
current performance.

The key difference is that existing OCO-with-memory formulations usually take a
finite memory length as part of the problem definition and do not make sparse
communication a source of trajectory mismatch. In contrast, \textsc{OCO-S$^2$}
starts from a stable dynamical system: the current state, and therefore the
incurred loss, depends in principle on the entire past decision history.
Moreover, the learner receives first-order feedback only through block-level
partial communication, so the deployed decision is held fixed within each
communication block while the state continues to evolve at the per-round time
scale. Thus sparse communication is not merely a feedback constraint; it also
changes the realized state trajectory on which future costs are incurred. We
exploit the fading-memory property of stable dynamics to control the
contribution of remote history and approximate the incurred trajectory cost by
finite-memory surrogate losses. In this sense, \textsc{OCO-S$^2$} extends
finite-memory OCO from explicitly windowed losses with regular feedback to
stateful incurred costs under sparse block communication.

\paragraph{Sparse communication and partial participation.}
Communication constraints are central in distributed online learning and
communication-efficient optimization. Existing methods study how infrequent
synchronization, incomplete feedback, heterogeneous local information, or
partial participation affect optimization and regret performance
\cite{stich2019localsgd,karimireddy2020scaffold,
malinovskiy2020localfixedpoint,jhunjhunwala2022fedvarp,
lin2022decentralizedoco,wan2022projectionfreedistributed,
cao2023compresseddecentralizedoco}. In pointwise loss models, sparse
communication mainly affects the update error or the variance of gradient
estimates. In \textsc{OCO-S$^2$}, sparse communication has an additional
effect: because decisions are held fixed within communication blocks, it
changes the realized state trajectory and therefore the future incurred costs.
Moreover, the online learner updates only at block boundaries, whereas the
dynamic comparator may vary at the per-round time scale under a path-length
budget.

Overall, existing work addresses memory-based online losses and
communication-constrained feedback from complementary perspectives.
\textsc{OCO-S$^2$} combines these elements in a single setting: losses are
induced by a stable dynamical system, communication occurs only at block
boundaries, first-order feedback is obtained under partial participation, and
performance is measured on the incurred trajectory cost.
\section{\textsc{OCO-S$^2$} Problem Setup}
\begin{figure}[t]
\centering
\includegraphics[width=\linewidth]{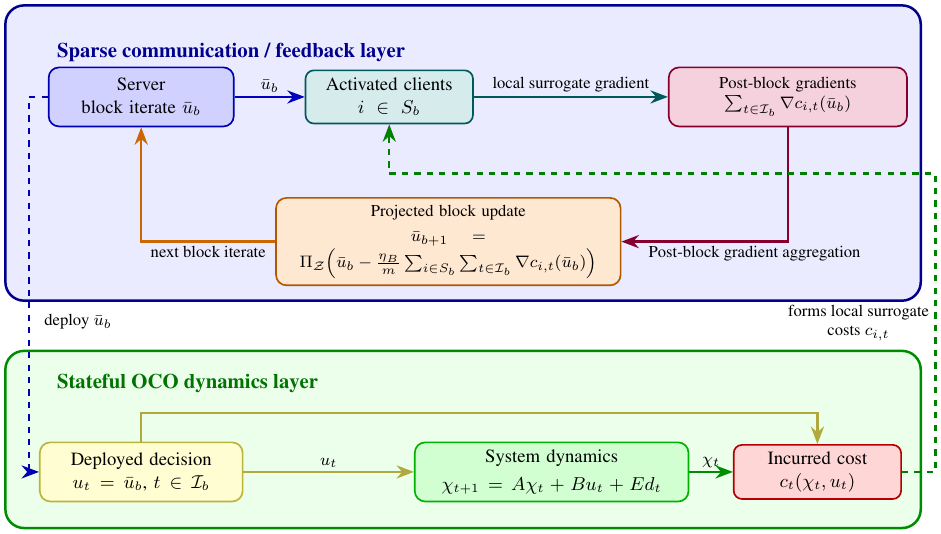}
\caption{\textsc{OCO-S$^2$}: separation between the stateful dynamics layer and the sparse communication layer. The learner holds a block decision, the state evolves at every round, and activated clients provide partial block-level first-order feedback.}
\label{fig:oco-s2-layer-separation}
\end{figure}
Before stating the \textsc{OCO-S$^2$} problem, we first fix the notation used throughout the paper.

\subsection{Notation}

For a finite set \(\mathcal S\), we use \(|\mathcal S|\) to denote its
cardinality. For two sets \(\mathcal A\) and \(\mathcal B\),
\(\mathcal A\times\mathcal B\) denotes their Cartesian product.
Throughout the paper, \(\|\cdot\|\) denotes the Euclidean norm for
vectors and the induced operator norm for matrices unless otherwise specified.
For any \(x\in\mathbb R\), \(\lceil x\rceil\) denotes the smallest integer no less than \(x\).
For a time-indexed quantity \(x_t\), and for any natural numbers \(a,b\) with
\(a\le b\), we write \(x_{a:b}:=(x_a,x_{a+1},\dots,x_b)\) for the trajectory
segment from time \(a\) to time \(b\). For integers \(a\) and \(b\), we write \(a\mid b\) if \(a\) divides \(b\), i.e., there exists an integer \(q\) such that \(b=qa\).
We use standard asymptotic notation. For nonnegative functions \(f\) and \(g\),
we write \(f=O(g)\) if there exist constants \(C>0\) and \(T_0>0\) such that
\(f\le Cg\) for all \(T\ge T_0\), and \(f=o(g)\) if \(f/g\to 0\) as
\(T\to\infty\).

\subsection{Stateful Online Decision Model}

We consider a communication-constrained online decision system consisting of a
central learner and \(N\) distributed clients that provide partial first-order
feedback. As illustrated in Fig.~\ref{fig:oco-s2-layer-separation}, the
operational system dynamics are separated from the sparse communication and
feedback layer. The dynamics below describe the operational system whose
behavior is affected by the deployed decisions.
Over the time horizon \(t=1,\dots,T\), the system state evolves according to the
discrete-time dynamics
\begin{equation}
\chi_{t+1}=A\chi_t+Bu_t+Ed_t,\quad \chi_0=0.
\label{eq:stateful-dynamics}
\end{equation}
The state \(\chi_t\in\mathbb R^{n_\chi}\), action
\(u_t\in\mathbb R^{n_u}\), and external input
\(d_t\in\mathbb R^{n_d}\) are system variables, where
\(n_\chi\), \(n_u\), and \(n_d\) denote the dimensions of the state, action, and
external-input spaces, respectively.

The comparator sequence introduced later serves as a hindsight benchmark for
dynamic-regret evaluation. It is propagated through the same
operational dynamics as the online sequence:
\begin{equation}
\chi_{t+1}^\star=A\chi_t^\star+B u_t^\star+E d_t,\quad \chi_0^\star=0.
\label{eq:comparator-dynamics}
\end{equation}
Whenever the dynamics are unrolled, we use the zero-padding convention
\(u_t=u_t^\star=0\) and \(d_t=0\) for all \(t\le 0\), so that
\(\chi_1=0\) and \(\chi_1^\star=0\). The online decision at round \(t\) is \(u_t\in\mathbb R^{n_u}\), and the
decision set is \(\mathcal Z\subset\mathbb R^{n_u}\). The algorithm deploys
decisions in \(\mathcal Z\) throughout the horizon. Likewise, the comparator
sequence satisfies \(u_t^\star\in\mathcal Z\) for every \(t\).
\subsection{Sparse Communication and Partial Participation}

Full synchronization at every round can incur prohibitive communication overhead
in practical communication-constrained online decision systems. We therefore
adopt a block-based periodic communication mechanism, rather than performing
full synchronization at every round.
Given a block length \(K\in\{1,\dots,T\}\), we partition the time horizon into
\(B_T:=\left\lceil T/K\right\rceil\) blocks and denote the time-index set of block \(b\)
by
\begin{equation}
\mathcal I_b
:=
\{(b-1)K+1,\dots,\min(bK,T)\},
\quad b=1,\dots,B_T.
\label{eq:block-index-set}
\end{equation}

Within each block, the server communicates with the clients only once at the
block boundary. Moreover, each communication round activates only \(m\) clients
rather than all \(N\) clients. Let the activated client set in block \(b\) be
\(S_b\subseteq\{1,\dots,N\}\), with \(|S_b|=m\). The sampling rule used in the
analysis is specified in Assumption~\ref{ass:partial_participation}.
Thus, \(K\) controls the communication frequency, while \(m\) controls the scale
per communication round.

\subsection{Dynamic-Regret Metric}

We evaluate decisions through a stage cost that depends jointly on the current
state and the current decision, denoted by \(c_t(\chi_t,u_t)\). Given an online decision sequence
\(u_{1:T}\), its incurred cumulative cost is defined as
\begin{equation}
J_T(u_{1:T}):=\sum_{t=1}^T c_t(\chi_t,u_t).
\label{eq:incurred-cumulative-cost}
\end{equation}
For a comparator sequence \(u_{1:T}^\star\), with the induced state
trajectory defined by \eqref{eq:comparator-dynamics}, its cumulative cost is
defined as
\begin{equation}
J_T(u_{1:T}^\star)
:=
\sum_{t=1}^T c_t(\chi_t^\star,u_t^\star).
\label{eq:comparator-cumulative-cost}
\end{equation}
To obtain a benchmark that remains meaningful under nonstationary disturbances,
we compare against a path-length-bounded dynamic comparator class
\begin{equation}
\mathcal C(V_T)
=
\left\{
u_{1:T}^\star\in\mathcal Z^T:
\sum_{t=2}^T \|u_t^\star-u_{t-1}^\star\|\le V_T
\right\},
\label{eq:dynamic-comparator-class}
\end{equation}
where \(V_T\ge0\) measures the total variation of the comparator decision
trajectory. The case \(V_T=0\) recovers the best fixed
decision, while larger \(V_T\) allows the benchmark trajectory to track
environmental drift.

The performance metric of interest is the dynamic regret
\begin{equation}
\mathrm{Reg}_{T}(V_T)
:=
J_T(u_{1:T})
-
\min_{u_{1:T}^\star\in\mathcal C(V_T)}J_T(u_{1:T}^\star).
\label{eq:dynamic-regret-metric}
\end{equation}
\subsection{Analysis Assumptions}
\label{subsec:analysis_assumptions}

Building on the stateful decision model, sparse communication protocol, and
dynamic-regret metric defined above, we impose the following assumptions for the
regret analysis. These assumptions also support the finite-memory surrogate
construction introduced in the next subsection.

\begin{assumption}[Action domain]
\label{ass:action_domain}
The action domain \(\mathcal Z\) is nonempty, convex, and compact. Consequently,
the decision radius and the diameter of \(\mathcal Z\) are uniformly bounded. We
define $R:=\max_{u\in\mathcal Z}\|u\|, D:=\max_{u,u'\in\mathcal Z}\|u-u'\|.$
\end{assumption}

\begin{assumption}[Bounded inputs]
\label{ass:bounded_inputs}
There exists a constant \(D_d>0\) such that $\|d_t\|\le D_d, \forall t=1,\dots,T.$
\end{assumption}

\begin{assumption}[Fading memory]
\label{ass:stable_dynamics}
There exist constants \(C_A\ge 1\) and \(\rho\in(0,1)\) such that $\|A^k\|\le C_A\rho^k,\forall k\ge 0.$
\end{assumption}

\begin{assumption}[Partial participation]
\label{ass:partial_participation}
At each post-block communication round, the active client set \(S_b\) is
sampled uniformly without replacement from all size-\(m\) subsets of
\(\{1,\dots,N\}\), conditional on the information available after the current
block is realized and before this sampling step.
\end{assumption}

\begin{assumption}[Local surrogate decomposition]
\label{ass:local_surrogate_decomposition}
For the diagonal surrogate state \(\bar\chi_t(u)\) introduced in the next
subsection, the corresponding surrogate cost admits the decomposition
\begin{equation}
c_t(\bar\chi_t(u),u)
=
\frac{1}{N}\sum_{i=1}^N c_{i,t}(u),
\label{eq:local-surrogate-decomposition}
\end{equation}
where each local stationary surrogate cost \(c_{i,t}\) is convex and
continuously differentiable on \(\mathcal Z\).
\end{assumption}

\begin{assumption}[Bounded gradients and smoothness]
\label{ass:bounded_gradients_smoothness}
There exist constants \(G_{\mathrm{loc}}, G_{\beta}, L_{\beta}>0\) such that
for any \(i\in\{1,\dots,N\}\), any \(t=1,\dots,T\), and any
\(u,u'\in\mathcal Z\),
\begin{equation}
\|\nabla c_{i,t}(u)\|\le G_{\mathrm{loc}},
\qquad
\left\|\frac{1}{N}\sum_{j=1}^N \nabla c_{j,t}(u)\right\|\le G_{\beta},
\end{equation}
and
\begin{equation}
\|\nabla c_{i,t}(u)-\nabla c_{i,t}(u')\|
\le
L_{\beta}\|u-u'\|.
\end{equation}
\end{assumption}

\begin{assumption}[Gradient heterogeneity]
\label{ass:gradient_heterogeneity}
There exists a constant \(\sigma_{\mathrm{het}}^2\ge 0\) such that for any
\(t=1,\dots,T\) and any \(u\in\mathcal Z\),
\begin{equation}
\frac{1}{N}\sum_{i=1}^N
\left\|
\nabla c_{i,t}(u)
-
\frac{1}{N}\sum_{j=1}^N\nabla c_{j,t}(u)
\right\|^2
\le
\sigma_{\mathrm{het}}^2.
\end{equation}
\end{assumption}

\begin{assumption}[State-Lipschitz real cost]
\label{ass:state_lipschitz_real_cost}
There exists a uniformly bounded set \(X_{\mathrm{all}}\) that contains all
true states generated by feasible decision sequences
\(u_{1:T}\in \mathcal Z\times\cdots\times\mathcal Z\), all finite-memory
truncated states, and all diagonal surrogate states. Moreover, there exists a
constant \(L_{\chi}>0\) such that for any \(u\in\mathcal Z\), any
\(\chi,\chi'\in X_{\mathrm{all}}\), and any \(t=1,\dots,T\),
\begin{equation}
\bigl|c_t(\chi,u)-c_t(\chi',u)\bigr|
\le
L_{\chi}\|\chi-\chi'\|.
\end{equation}
\end{assumption}

Compactness of \(\mathcal Z\) ensures uniformly bounded decisions, bounded
inputs rule out unbounded disturbances, and the stability condition on \(A\)
gives a fading-memory property of the state dynamics. Partial participation and
the local surrogate decomposition define the distributed feedback available to
the learner, while the gradient and state-Lipschitz conditions control the
surrogate and incurred-cost errors used in the regret analysis.
\subsection{Finite-Memory Surrogate Losses}

The main analytical difficulty comes from the state recursion
\eqref{eq:stateful-dynamics}. Once the dynamics are unrolled, the current state
depends on the entire history of past actions and external inputs. Thus, the
incurred cost \(c_t(\chi_t,u_t)\) is not a pointwise loss in the current decision
alone. Nevertheless, by Assumption~\ref{ass:stable_dynamics}, the effect of
remote history decays geometrically through the powers of \(A\). This
fading-memory property allows us to approximate the realized state by a
finite-window state, with the truncation error controlled by the stability of
the dynamics.

Under the zero-padding convention introduced above, let \(H\) be a positive
integer that specifies the memory length.
Along the algorithm trajectory, we define the finite-window state
\begin{equation}
\hat\chi_t(u_{t-H:t})
:=
\sum_{i=0}^{H-1}A^i(Bu_{t-1-i}+Ed_{t-1-i}).
\label{eq:finite-window-state}
\end{equation}
This quantity keeps only the most recent \(H\) rounds of actions and external
inputs, while the contribution from earlier rounds is controlled by
Assumption~\ref{ass:stable_dynamics}.

However, the finite-window state \(\hat\chi_t(u_{t-H:t})\) still depends on all
action inputs inside the window. To obtain a pointwise surrogate that can be
optimized using standard online-gradient updates, we further define, for any
single decision \(u\), the diagonal surrogate state
\begin{equation}
\bar\chi_t(u)
:=
\sum_{i=0}^{H-1}A^i(Bu+Ed_{t-1-i}).
\label{eq:diagonal-surrogate-state}
\end{equation}
Compared with \(\hat\chi_t\), this diagonal state replaces all action inputs in
the window by the same current candidate action \(u\), while keeping the
observed disturbance window unchanged.

We compare the realized finite-window cost
\(c_t(\hat\chi_t(u_{t-H:t}),u_t)\) with the corresponding diagonal surrogate
cost \(c_t(\bar\chi_t(u),u)\). The local decomposition and regularity
conditions needed to form distributed block gradients are specified in
Assumptions~\ref{ass:local_surrogate_decomposition}--\ref{ass:state_lipschitz_real_cost}.
\section{Algorithm}
Algorithm~\ref{alg:block_periodic_update} implements this block-level
partial-participation communication model using the local stationary surrogate
costs \(c_{i,t}\). The computation of these local surrogate costs is
model-based: the system matrices \(A,B,E\) in \eqref{eq:stateful-dynamics} are
assumed known, and after block \(\mathcal I_b\) is realized, the required
disturbance window is available. Hence each activated client can construct, or
equivalently query from the surrogate-gradient layer, its local component
\(c_{i,t}\) induced by the diagonal state \(\bar\chi_t(u)\), and evaluate
\(\nabla c_{i,t}(\bar u_b)\). At the beginning of block \(b\), the server
deploys \(\bar u_b\), so \(u_t=\bar u_b\) for all \(t\in\mathcal I_b\). After
the block is observed, the activated clients return local block gradients at
\(\bar u_b\), and the projected update produces \(\bar u_{b+1}\) for the next
block.
\begin{algorithm}[t]
\caption{\textsc{OCO-S$^2$-OGD}: Block Online Gradient Descent with Sparse Communication}
\label{alg:block_periodic_update}
\begin{algorithmic}[1]
\Require Block length \(K\), memory length \(H\), step size \(\eta_B\), number of activated clients per communication round \(m\), initial point \(\bar u_1\in\mathcal Z\)
\For{\(b=1,2,\dots,B_T\)}
\State For all \(t\in\mathcal I_b\), set \(u_t \gets \bar u_b\).
\State The operational system evolves over block \(\mathcal I_b\); the corresponding losses and disturbance window are observed, and local surrogate-gradient information is constructed using the known system model.
\State Uniformly sample a client subset \(S_b\subseteq\{1,\dots,N\}\) without replacement for the post-block communication round such that \(|S_b|=m\).
\State Make \(\bar u_b\) available to each activated client \(i\in S_b\).
\State After block \(\mathcal I_b\) has been observed, each activated client \(i\in S_b\) evaluates and returns the post-block gradient \(\sum_{t\in\mathcal I_b}\nabla c_{i,t}\bigl(\bar u_b\bigr)\).
\State Update
\begin{equation}
\begin{aligned}
\bar u_{b+1}
\gets
\Pi_{\mathcal Z}\left(
\bar u_b
- \frac{\eta_B}{m}
\sum_{i\in S_b}
\sum_{t\in\mathcal I_b}
\nabla c_{i,t}\bigl(\bar u_b\bigr)
\right).
\end{aligned}
\label{eq:oco-s2-ogd-block-update}
\end{equation}
\EndFor
\Ensure Online decision sequence \(u_{1:T}\)
\end{algorithmic}
\end{algorithm}
\section{Main Results}
The main results quantify the communication cost of \textsc{OCO-S$^2$-OGD} and
its dynamic-regret guarantee for the incurred state-dependent cost under the
assumptions stated in Section~\ref{subsec:analysis_assumptions}.
\subsection{Communication Complexity}

We quantify the algorithmic synchronization cost of the block protocol. In this
paper, communication complexity refers only to the number of scalars exchanged
for block-level synchronization between the server and the activated clients:
the block iterate made available to each activated client and the returned
gradient vector. It does not include implementation-dependent operational
feedback, broadcasting mechanisms, or other system-layer information flows,
which may vary across deployments.

\begin{proposition}[Synchronization complexity]
\label{prop:comm_complexity}
Under the above convention, the synchronization cost of
Algorithm~\ref{alg:block_periodic_update} satisfies
\begin{equation}
\mathrm{Comm}_{\mathrm{blk}}(T,K)
=
2m n_u\left\lceil \frac{T}{K}\right\rceil.
\label{eq:block-communication-total}
\end{equation}
In particular, when
\(K=\sqrt{T}\), we obtain
\begin{equation}
\mathrm{Comm}_{\mathrm{blk}}(T,\sqrt{T})=O(\sqrt{T}).
\label{eq:block-communication-sqrt}
\end{equation}
\end{proposition}

The dependence on \(K\) makes the synchronization--performance tradeoff
explicit: larger blocks reduce the number of synchronization rounds, while the
regret theorem below shows how this choice enters the performance guarantee.
\subsection{Dynamic Regret for the Incurred Cost}

The following result shows that \textsc{OCO-S$^2$-OGD} retains dynamic-regret
control for \textsc{OCO-S$^2$} under sparse block communication and partial
participation.

\begin{theorem}[Dynamic regret under sparse block communication]
\label{thm:real_dyn_regret_informal}
Run Algorithm~\ref{alg:block_periodic_update} with initialization
\(\bar u_1\in\mathcal Z\), block length \(K\), memory length \(H\), and block step
size \(\eta_B\). Then there exist constants
\(C_0,C_1,C_2,C_3,C_4,C_5>0\), independent of
\(T\), \(K\), \(H\), \(V_T\), and \(\eta_B\), such that
\begin{equation}
\begin{aligned}
\mathbb E\!\left[\mathrm{Reg}_{T}(V_T)\right]
\le{}&
\frac{C_0}{\eta_B}
+
\frac{C_1V_T}{\eta_B}
+
C_2\eta_BKT\\
&+
C_3KV_T
+
C_4V_T
+
C_5T\rho^H .
\end{aligned}
\label{eq:main-regret-bound-general}
\end{equation}
In particular, if $\eta_B=\frac{1}{\sqrt{KT}}, H=\left\lceil \frac{\log T}{|\log \rho|}\right\rceil$
then
\begin{equation}
\begin{aligned}
\mathbb E\!\left[\mathrm{Reg}_{T}(V_T)\right]
\le{}&
(C_0+C_2)\sqrt{KT}
+
C_1V_T\sqrt{KT}\\
&+
C_3KV_T
+
C_4V_T
+
C_5 .
\end{aligned}
\label{eq:main-regret-bound-tuned}
\end{equation}
When \(K=\left\lceil\sqrt{T}\right\rceil\), Proposition~\ref{prop:comm_complexity}
gives \(O(\sqrt{T})\) communication and
\begin{equation}
\begin{aligned}
\mathbb E\!\left[\mathrm{Reg}_{T}(V_T)\right]
\le{}&
(C_0+C_2)T^{3/4}
+
C_1V_TT^{3/4}\\
&+
C_3V_TT^{1/2}
+
C_4V_T
+
C_5 .
\end{aligned}
\label{eq:main-regret-bound-sqrt}
\end{equation}
\end{theorem}
\paragraph{Discussion and limiting cases.}
Theorem~\ref{thm:real_dyn_regret_informal} should be read as a decomposition
of the additional prices paid by \textsc{OCO-S$^2$-OGD} beyond standard online
gradient descent. The \(C_2\eta_BKT\) term is the block-level optimization
price, the \(C_3KV_T\) and \(C_4V_T\) terms capture comparator drift and
block alignment, and \(C_5T\rho^H\) is the finite-memory truncation term.
This decomposition makes several limiting cases transparent.

First, when \(V_T=0\), the comparator class reduces to the best fixed
comparator in hindsight. In this static-comparator case, the bound becomes
\begin{equation}
\mathbb E[\mathrm{Reg}_T(0)]
=
O\!\left(\frac{1}{\eta_B}\right)
+
O(\eta_BKT)
+
O(T\rho^H),
\end{equation}
and under the tuned parameters,
\begin{equation}
\mathbb E[\mathrm{Reg}_T(0)]
=
O(\sqrt{KT})+O(1).
\end{equation}
Thus, with \(K=\lceil\sqrt T\rceil\), \textsc{OCO-S$^2$-OGD} obtains
\(O(T^{3/4})\) regret with \(O(\sqrt T)\) communication, whereas with
per-round communication \(K=1\), it recovers the familiar
\(O(\sqrt T)\) static-regret rate up to the controlled memory-truncation error.

Second, when block communication is removed by setting \(K=1\), the learner is
allowed to update at every round. The tuned bound reduces to
\begin{equation}
\mathbb E[\mathrm{Reg}_T(V_T)]
=
O\!\left((1+V_T)\sqrt T\right)
+
O(V_T)
+
O(1),
\end{equation}
which matches the standard dynamic-regret scaling of online gradient methods
up to the additional state-memory approximation term. In particular, for any
fixed variation budget \(V_T=O(1)\), the regret is \(O(\sqrt T)\). More
generally, in the per-round communication regime, sublinearity is preserved
whenever \(V_T=o(\sqrt T)\).

Third, when partial participation is removed by taking \(m=N\), the block
gradient estimator becomes the full distributed gradient of the surrogate block
loss. In this case, the sampling error caused by partial participation
vanishes. The asymptotic dependence on \(T\), \(K\), \(H\), and \(V_T\) remains
the same, but the constants associated with participation-induced gradient
noise and client heterogeneity disappear. Thus full participation improves the
constant factors but does not remove the fundamental block-alignment and
state-memory terms, which are caused by sparse communication and stateful
incurred costs rather than by sampling.

Fourth, if the state dependence is exactly finite-memory, or if \(H\) is chosen
large enough so that the truncation tail is negligible, the term
\(T\rho^H\) disappears from the leading-order bound. In the further special
case of pointwise losses, where the current loss depends only on the current
decision, \textsc{OCO-S$^2$} reduces to a block-feedback OCO problem. The bound
then keeps only the online-gradient, comparator-drift, and block-alignment
terms. With \(K=1\) and \(m=N\), this further reduces to the usual full-feedback
OCO regime with dynamic regret
\begin{equation}
O\!\left((1+V_T)\sqrt T\right).
\end{equation}
Hence the theorem recovers the standard OCO behavior in the absence of sparse
communication and state memory, while quantifying precisely how the additional
\textsc{OCO-S$^2$} features modify the regret.
\subsection{Benefiting from Predictions}
\label{subsec:benefiting_from_predictions}

We next consider a prediction-augmented variant of
\textsc{OCO-S$^2$-OGD}. Before block \(b\) is played, the learner receives an
exogenous block-level prediction \(M_b\in\mathbb R^{n_u}\) satisfying
\begin{equation}
\|M_b\|\le K G_P .
\end{equation}
The prediction is interpreted as advice for the full surrogate gradient over
the upcoming block. We do not model how \(M_b\) is generated; it may come from
forecasting, historical data, or any external prediction module. The resulting
algorithm, \textsc{OCO-S$^2$-OGD-P}, is given in
Algorithm~\ref{alg:oco_s2_ogd_p}.

We measure prediction quality by the cumulative mismatch between \(M_b\) and
the realized full block-level surrogate gradient:
\begin{equation}
\mathcal E_T(M)
:=
\sum_{b=1}^{B_T}
\mathbb E
\left\|
\frac1N
\sum_{i=1}^N
\sum_{t\in\mathcal I_b}
\nabla c_{i,t}(\bar u_b)
-
M_b
\right\|^2.
\end{equation}
\begin{theorem}[Dynamic regret with predictions]
\label{thm:prediction_augmented_regret}
Run \textsc{OCO-S$^2$-OGD-P} with initialization \(q_1\in\mathcal Z\), block
length \(K\), memory length \(H\), block step size \(\eta_B\), and prediction
sequence \(\{M_b\}_{b=1}^{B_T}\) satisfying \(\|M_b\|\le KG_P\). Then there
exist constants
\(C_0,C_1,C_2^{\mathrm{res}},C_3,C_4,C_5,C_6>0\), independent of
\(T\), \(K\), \(H\), \(V_T\), \(\eta_B\), and the prediction sequence except
through \(\mathcal E_T(M)\), such that
\begin{equation}
\begin{aligned}
\mathbb E[\mathrm{Reg}_T(V_T)]
\le{}&
\frac{C_0}{\eta_B}
+
\frac{C_1V_T}{\eta_B}\\
&+
\eta_B\!\left(C_2^{\mathrm{res}}KT+C_6\mathcal E_T(M)\right)\\
&+
C_3KV_T
+
C_4V_T
+
C_5T\rho^H .
\end{aligned}
\end{equation}
In particular, if $\eta_B
=
\frac{1}{\sqrt{KT}}, H=
\left\lceil\frac{\log T}{|\log\rho|}\right\rceil,$
then
\begin{equation}
\begin{aligned}
\mathbb E[\mathrm{Reg}_T(V_T)]
\le{}&
\left(C_0+C_2^{\mathrm{res}}\right)\sqrt{KT}
+
\frac{C_6\mathcal E_T(M)}{\sqrt{KT}}\\
&+
C_1V_T\sqrt{KT}
+
C_3KV_T
+
C_4V_T
+
C_5 .
\end{aligned}
\end{equation}
\end{theorem}
Compared with Theorem~\ref{thm:real_dyn_regret_informal}, the prediction
variant changes only the optimization term. In the baseline bound, this term is
\(C_2\eta_BKT\). In the prediction-augmented bound, it becomes
\begin{equation}
\eta_B\!\left(C_2^{\mathrm{res}}KT+C_6\mathcal E_T(M)\right),
\end{equation}
where \(C_2^{\mathrm{res}}\) collects the prediction-insensitive residual
effects and \(C_6\mathcal E_T(M)\) measures the realized prediction mismatch.

Thus predictions are beneficial when the mismatch is small enough that
\begin{equation}
C_2^{\mathrm{res}}KT+C_6\mathcal E_T(M)
<
C_2KT .
\end{equation}
Under the fixed baseline stepsize \(\eta_B=1/\sqrt{KT}\), this is exactly the
condition under which the prediction-augmented optimization term is smaller
than the baseline optimization term. The remaining terms---comparator drift,
block alignment, state-memory truncation, and partial-participation effects
outside the optimization coefficient---are unchanged. In the extreme case
\(M_b=0\), \textsc{OCO-S$^2$-OGD-P} reduces to the baseline
\textsc{OCO-S$^2$-OGD}; under the bounded-gradient assumptions,
\(\mathcal E_T(M)=O(KT)\), so the prediction-augmented bound recovers the
baseline order up to constants.
\begin{algorithm}[t]
\caption{\textsc{OCO-S$^2$-OGD-P}: Prediction-Augmented Block Online Gradient Descent}
\label{alg:oco_s2_ogd_p}
\begin{algorithmic}[1]
\Require Block length \(K\), memory length \(H\), step size \(\eta_B\), number of activated clients per communication round \(m\), initial anchor \(q_1\in\mathcal Z\)
\For{\(b=1,2,\dots,B_T\)}
\State Receive an exogenous prediction vector \(M_b\in\mathbb R^{n_u}\) with \(\|M_b\|\le K G_P\).
\State Form the optimistic block decision
\begin{equation}
\bar u_b\gets \Pi_{\mathcal Z}(q_b-\eta_B M_b).
\end{equation}
\State For all \(t\in\mathcal I_b\), set \(u_t\gets \bar u_b\).
\State The operational system evolves over block \(\mathcal I_b\); the corresponding losses and disturbance window are observed, and local surrogate-gradient information is constructed using the known system model.
\State Uniformly sample a client subset \(S_b\subseteq\{1,\dots,N\}\) without replacement such that \(|S_b|=m\).
\State Each activated client \(i\in S_b\) returns
\begin{equation}
\sum_{t\in\mathcal I_b}\nabla c_{i,t}(\bar u_b).
\end{equation}
\State Update the anchor
\begin{equation}
q_{b+1}
\gets
\Pi_{\mathcal Z}\left(
q_b
-
\frac{\eta_B}{m}
\sum_{i\in S_b}
\sum_{t\in\mathcal I_b}
\nabla c_{i,t}(\bar u_b)
\right).
\end{equation}
\EndFor
\Ensure Online decision sequence \(u_{1:T}\)
\end{algorithmic}
\end{algorithm}
\section{Conclusion}

We studied \textsc{OCO-S$^2$}, an online convex optimization problem with
stateful incurred costs under sparse block communication and partial
participation. Within this framework, \textsc{OCO-S$^2$-OGD} provides a
projected block online-gradient method and admits a dynamic-regret guarantee
for the incurred cost along the realized state trajectory. More broadly, this
work provides a regret-theoretic foundation for communication-efficient online
decision-making in systems where algorithmic updates and physical state
trajectories are intrinsically coupled.

\section*{Acknowledgments}
This work was supported in part by the National Natural Science Foundation of
China under Grant 52307145, and in part by the Shenzhen Research Institute of
Big Data (SRIBD) under Grant J00220250001.
\bibliography{reference}

\end{document}


\maketitle

\appendix
\section{Numerical Verification}

The numerical experiments provide mechanism-level verification of the regret
bound: the communication--regret tradeoff, finite-memory truncation, partial
participation, and horizon scaling.
\subsection{Controlled Synthetic \textsc{OCO-S$^2$} Instance}

We use a controlled synthetic \textsc{OCO-S$^2$} instance: a transparent stable
linear system with drifting disturbances, box-constrained online decisions, and
a path-length-bounded dynamic comparator.

The instance uses \(N=10\) clients, \(n_u=10\), and \(n_\chi=10\). The
dynamics are diagonal and stable, $A=0.95I, B=0.1I, E=-0.1I,$
and the decision set is the coordinate-wise box $\mathcal Z=\{u:0\le u\le 1\}.$
The disturbance sequence is generated synthetically from coordinate-wise
sinusoidal components, piecewise level shifts, and small clipped noise. The
stage cost is
\[
c_t(\chi_t,u_t)
=
\frac{\alpha}{2}\|\chi_t\|_2^2
+
\frac{\beta}{2}\|u_t-d_t\|_2^2,
\]
with \(\alpha=0.2\) and \(\beta=0.8\). The diagonal structure yields a simple
local surrogate decomposition: client \(i\) owns the \(i\)-th coordinate
contribution of \(c_t(\bar\chi_t(u),u)\), receives the broadcast vector
\(\bar u_b\), and returns the gradient of its assigned local surrogate
component. The offline comparator solves the convex conic problem over
\(u_{1:T}^\star\), subject to the same dynamics, the same box constraints, and
the Euclidean path budget in \eqref{eq:dynamic-comparator-class}. Unless a
sweep overrides the budget explicitly, we set
\[
V_T=0.45\sum_{t=2}^T\|d_t-d_{t-1}\|_2.
\]
We report final dynamic regret, average dynamic regret, and communication
scalars throughout.
To verify that the hindsight comparator is solved accurately, we also aggregate
offline-comparator diagnostics over the controlled sweeps used in the numerical
verification. \Cref{tab:comparator_diagnostics} reports the solve success rate,
realized path length, budget slack, and residual checks for the convex conic
comparator problems.

\begin{table*}[t]
\centering
\caption{Aggregated offline comparator diagnostics over the controlled synthetic sweeps used in the numerical verification.}
\label{tab:comparator_diagnostics}
\scriptsize
\setlength{\tabcolsep}{4pt}
\begin{tabular}{ccccc}
\toprule
Success rate & Solver & Iterations & \shortstack{Realized path length\\ mean [min, max]} & \shortstack{Budget slack\\ mean [min, max]} \\
\midrule
\ComparatorDiagnosticsSummaryRows
\bottomrule
\end{tabular}

\vspace{0.5em}

\begin{tabular}{cccc}
\toprule
\shortstack{Max dynamics\\ residual} & \shortstack{Max box-constraint\\ violation} & \shortstack{Max path-budget\\ violation} & \shortstack{Max relative objective\\ mismatch} \\
\midrule
\ComparatorDiagnosticsResidualRows
\bottomrule
\end{tabular}
\end{table*}

\subsection{Communication--Regret Tradeoff}

\begin{figure}[t]
\centering
\begin{subfigure}[t]{0.48\linewidth}
\centering
\includegraphics[width=\linewidth]{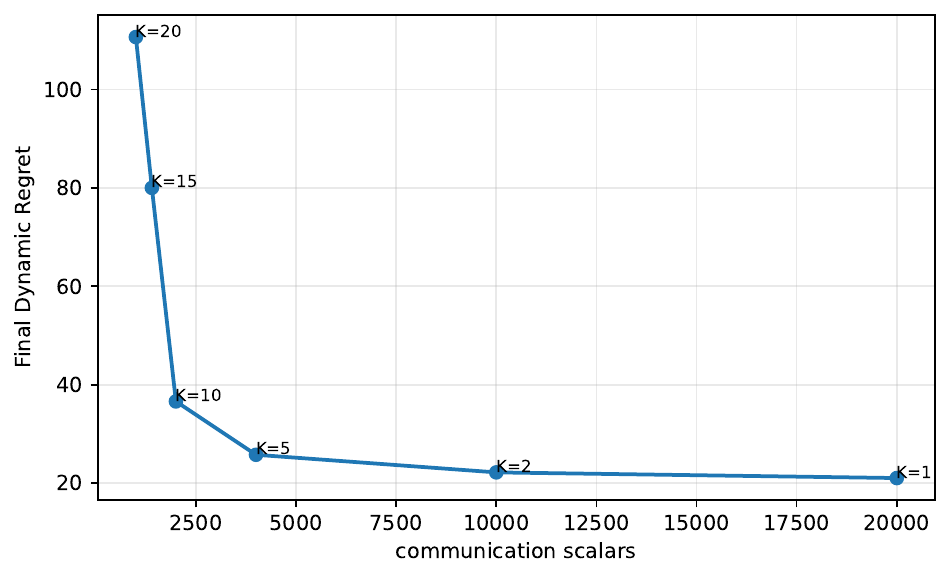}
\caption{Regret vs. communication.}
\label{fig:k_tradeoff_regret_vs_comm}
\end{subfigure}\hfill
\begin{subfigure}[t]{0.48\linewidth}
\centering
\includegraphics[width=\linewidth]{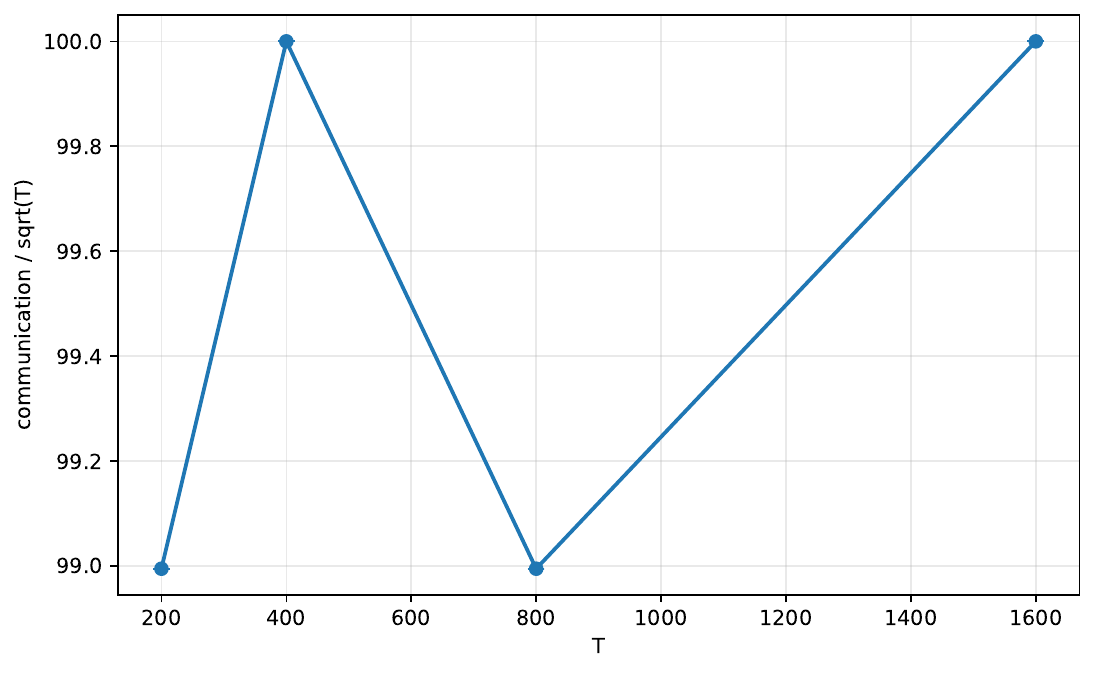}
\caption{Communication normalized by \(\sqrt T\).}
\label{fig:scaling_comm_over_sqrtt}
\end{subfigure}
\caption{Synchronization-sensitive sweep evidence on the controlled synthetic \textsc{OCO-S$^2$} instance.}
\label{fig:core_results}
\end{figure}

We first examine the synchronization effect of the block protocol. We vary the
block length \(K\) while holding the remaining parameters fixed. Larger blocks
reduce synchronization because the number of communication rounds decreases,
but they also increase the regret terms associated with block-held decisions
and delayed correction. \Cref{fig:k_tradeoff_regret_vs_comm} shows this
synchronization--regret tradeoff directly.

We also vary the horizon length \(T\) under the operating point
\(K=\lceil\sqrt T\rceil\). \Cref{fig:scaling_comm_over_sqrtt} reports the
normalized synchronization cost, which remains stable as \(T\) increases and
matches $\mathrm{Comm}_{\mathrm{blk}}\!\left(T,\left\lceil \sqrt{T}\right\rceil\right)
=O(\sqrt{T}).$
Thus, the experiments corroborate the block-length communication--regret
tradeoff and the predicted synchronization scaling under
\(K=\lceil\sqrt T\rceil\).

\subsection{Effect of Memory Length}

We next sweep the memory length \(H\). Increasing \(H\) improves the
finite-memory approximation of the stateful cost until the truncation error
becomes negligible, after which the curve saturates. This behavior is
consistent with the \(T\rho^H\) term in the regret bound.

\subsection{Effect of Partial Participation}

\begin{figure}[t]
\centering
\begin{subfigure}[t]{0.48\linewidth}
\centering
\includegraphics[width=\linewidth]{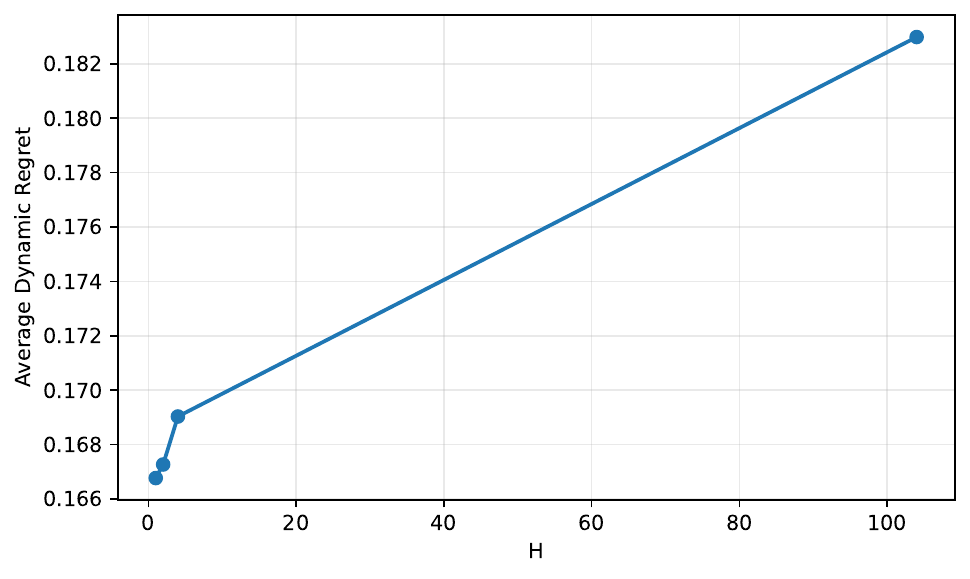}
\caption{Average dynamic regret versus memory length \(H\).}
\label{fig:memory_effect}
\end{subfigure}\hfill
\begin{subfigure}[t]{0.48\linewidth}
\centering
\includegraphics[width=\linewidth]{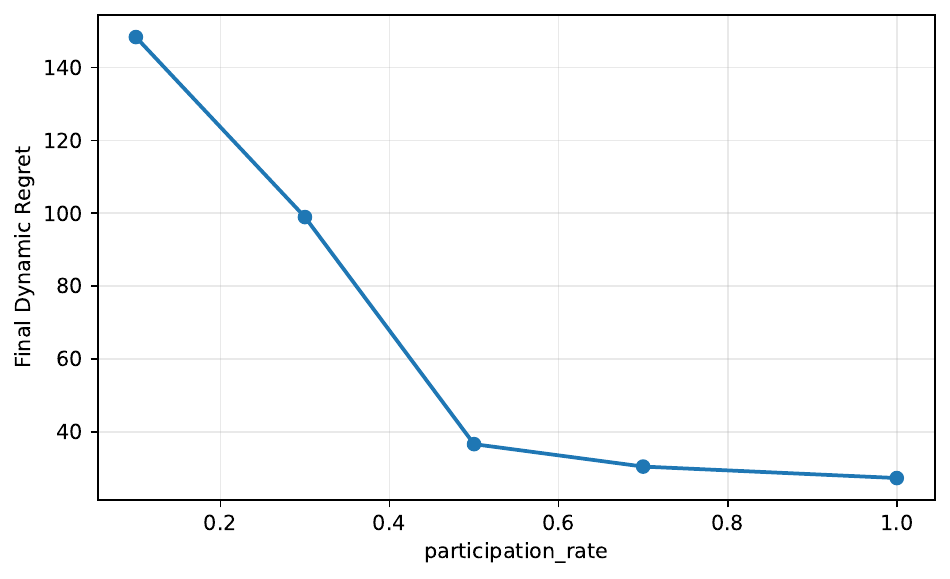}
\caption{Final dynamic regret under partial participation.}
\label{fig:participation_effect}
\end{subfigure}
\caption{Memory and participation effects on the controlled synthetic \textsc{OCO-S$^2$} instance.}
\label{fig:memory_participation_effects}
\end{figure}

We then vary the number of activated clients through the participation ratio
\(m/N\). Lower participation reduces the gradients incorporated in each block
update and therefore increases sampling noise. \Cref{fig:participation_effect}
shows the corresponding increase in incurred-cost dynamic regret as \(m\)
decreases.

\section{Additional Experimental Details}

This appendix collects implementation details for the controlled synthetic
instance. Table~\ref{tab:appendix-setup-details} summarizes the default
instance, algorithm, and comparator settings used throughout the numerical
verification.

\begin{table}[t]
\centering
\caption{Additional setup details for the controlled synthetic \textsc{OCO-S$^2$} instance.}
\label{tab:appendix-setup-details}
\small
\setlength{\tabcolsep}{3pt}
\begin{tabular}{>{\raggedright\arraybackslash}p{0.42\linewidth}
                >{\raggedright\arraybackslash}p{0.46\linewidth}}
\toprule
Item & Setting \\
\midrule
Number of clients & \(N=10\) \\
State/action dimensions & \(n_\chi=10\), \(n_u=10\) \\
Stable dynamics & \(A=0.95I\), \(B=0.1I\), \(E=-0.1I\) \\
Disturbance generator & Coordinate-wise sinusoids, piecewise shifts, and clipped noise \\
Box constraint & \(\mathcal Z=\{u:0\le u\le 1\}\) \\
Sweep horizon & \(T=200\) for the \(K\)-, \(H\)-, and participation sweeps \\
Scaling horizons & \(T\in\{200,400,800,1600\}\) \\
Random seeds & 5 per setting \\
Default block length & \(K=10\) \\
Default block step size & \(\eta_B=0.04\) \\
Default finite-memory horizon & Theory-sized \(H=\left\lceil \log T/|\log 0.95|\right\rceil\) \\
Default stage-cost weights & \(\alpha=0.2\), \(\beta=0.8\) \\
Default participation ratio & \(m/N=0.5\) (\(m=5\) out of \(N=10\)) \\
Default comparator budget & \(0.45\) times the disturbance Euclidean path length unless explicitly overridden \\
Local surrogate assignment & Client \(i\) owns the \(i\)-th coordinate contribution of the diagonal surrogate cost \\
Solver & CVXPY with CLARABEL \\
Reported uncertainty & Mean \(\pm\) one standard deviation when error bars are shown \\
\bottomrule
\end{tabular}
\end{table}

\paragraph{Synthetic local surrogate decomposition.}
In the synthetic instance, the diagonal surrogate cost is separable across
coordinates. Client \(i\) therefore owns the \(i\)-th coordinate contribution of
\(c_t(\bar\chi_t(u),u)\), receives the broadcast full vector \(\bar u_b\), and
computes the gradient of its assigned local surrogate component at that point.
The components need not share identical disturbance windows, and the feedback
may be supplied directly by the operational layer or relayed by the server.
\paragraph{Offline comparator computation and reported regret.}
For each disturbance realization, the offline comparator solves
\[
\begin{aligned}
\min_{u_{1:T}^\star}\quad
&\sum_{t=1}^T c_t(\chi_t^\star,u_t^\star)\\
\text{s.t.}\quad
&\chi_{t+1}^\star=A\chi_t^\star+B u_t^\star+E d_t,\\
&u_t^\star\in\mathcal Z,\qquad t=1,\dots,T,\\
&\sum_{t=2}^T \|u_t^\star-u_{t-1}^\star\|_2\le V_T.
\end{aligned}
\]
Unless a sweep overrides the budget explicitly, we set
\(V_T=0.45\sum_{t=2}^T \|d_t-d_{t-1}\|_2\). The offline comparator is solved as
a convex conic quadratic program with linear dynamics, box constraints, and a
Euclidean path-length budget. We solve it in CVXPY using
CLARABEL and independently recompute the objective and constraint residuals
from the returned trajectory. The aggregated diagnostics reported in
Table~\ref{tab:comparator_diagnostics} summarize these checks over the
controlled sweeps used in the numerical verification.
\section{Additional Proofs}
\subsection{Proof of Proposition~\ref{prop:comm_complexity}}

\begin{proof}
By the convention in Proposition~\ref{prop:comm_complexity}, we count only the
block-level synchronization exchange between the server and the activated
clients. Algorithm~\ref{alg:block_periodic_update} synchronizes once per block,
and the number of blocks is \(B_T=\lceil T/K\rceil\). In each synchronization
round, \(m\) clients are activated. For each activated client, the server makes
one \(n_u\)-dimensional block iterate available and receives one
\(n_u\)-dimensional gradient vector. Thus each activated client exchanges
\(2n_u\) scalars per block, and the total synchronization cost is
\[
\mathrm{Comm}_{\mathrm{blk}}(T,K)
=
2mn_u B_T
=
2mn_u\left\lceil \frac{T}{K}\right\rceil.
\]
For \(K=\sqrt T\), this becomes
\[
\mathrm{Comm}_{\mathrm{blk}}(T,\sqrt T)
=
2mn_u\left\lceil \sqrt T\right\rceil
=
O(\sqrt T).
\]
This proves the claim.
\end{proof}

\subsection{Formal Statement and Proof of the Dynamic-Regret Theorem}

\begin{theorem}[Formal dynamic-regret bound for the incurred cost]
\label{thm:real_dynamic_regret_formal}
Under Assumptions~\ref{ass:action_domain}--\ref{ass:state_lipschitz_real_cost}, 
define $C_{\mathrm{mem}}
:=
\frac{C_A\|B\|}{(1-\rho)^2},
D_\chi
:=
\frac{C_A}{1-\rho}\bigl(\|B\|R+\|E\|D_d\bigr),$
and $\Lambda_{\mathrm{blk}}
:=
\frac12\left(
G_\beta^2+\frac{N-m}{m(N-1)}\sigma_{\mathrm{het}}^2
\right)
+
L_\chi C_{\mathrm{mem}} G_{\mathrm{loc}}.$
Run Algorithm~\ref{alg:block_periodic_update} with initialization \(\bar u_1\in\mathcal Z\), block length \(K\), memory length \(H\), and block step size \(\eta_B\). Then, for every comparator sequence \(u_{1:T}^\star\in\mathcal C(V_T)\),
\[
\begin{aligned}
&\mathbb E\!\left[
J_T(u_{1:T})-J_T(u_{1:T}^\star)
\right]\\
&\le{}
\frac{D^2}{2\eta_B}
+
\frac{D}{\eta_B}V_T\\
&+
2\eta_BKT\Lambda_{\mathrm{blk}}
+
(G_\beta K+L_\chi C_{\mathrm{mem}})V_T\\
&+
2TL_\chi C_A D_\chi \rho^H.
\end{aligned}
\tag{A.1}
\label{eq:formal_dyn_regret_pre_tuning}
\]

In particular, if
\[
\eta_B=\frac{D}{2\sqrt{KT\Lambda_{\mathrm{blk}}}},
\qquad
H=\left\lceil \frac{\log T}{|\log \rho|}\right\rceil,
\]
then 
\[
\begin{aligned}
&\mathbb E\!\left[
J_T(u_{1:T})-J_T(u_{1:T}^\star)
\right]\\
&\le{}
2(D+V_T)\sqrt{KT\Lambda_{\mathrm{blk}}}\\
&+
(G_\beta K+L_\chi C_{\mathrm{mem}})V_T
+
2L_\chi C_A D_\chi.
\end{aligned}
\tag{A.2}
\label{eq:formal_dyn_regret_post_tuning}
\]
Consequently,
\[
\begin{aligned}
&\mathbb E\,\mathrm{Reg}_{T}(V_T)\\
&\le{}
2(D+V_T)\sqrt{KT\Lambda_{\mathrm{blk}}}\\
&+
(G_\beta K+L_\chi C_{\mathrm{mem}})V_T
+
2L_\chi C_A D_\chi.
\end{aligned}
\tag{A.3}
\label{eq:formal_dyn_regret_final}
\]
\end{theorem}

\begin{proof}
We prove the theorem in nine steps.
\paragraph{Step 1: Uniform state boundedness.}
We first record a uniform state bound used throughout the proof. By
the projected update in Algorithm~\ref{alg:block_periodic_update} and
Assumption~\ref{ass:action_domain}, the algorithm satisfies
\(\|u_t\|\le R\) for all \(t\). Unrolling
\eqref{eq:stateful-dynamics} gives
\[
\chi_t
=
\sum_{i=0}^{t-1}A^i(Bu_{t-1-i}+Ed_{t-1-i}).
\tag{A.4}
\label{eq:chi_expand_alg}
\]
Using Assumptions~\ref{ass:bounded_inputs} and \ref{ass:stable_dynamics}, we have
\[
\begin{aligned}
\|\chi_t\|
\le{}&
\sum_{i=0}^{t-1}C_A\rho^i(\|B\|R+\|E\|D_d)
\\
&\le
\frac{C_A}{1-\rho}(\|B\|R+\|E\|D_d)
=
D_\chi .
\end{aligned}
\tag{A.5}
\label{eq:Dchi_bound_alg}
\]

The same bound holds for every comparator
\(u_{1:T}^\star\in\mathcal C(V_T)\), since \(u_t^\star\in\mathcal Z\) implies
\(\|u_t^\star\|\le R\). Indeed,
\[
\chi_t^\star
=
\sum_{i=0}^{t-1}A^i(Bu_{t-1-i}^\star+Ed_{t-1-i}),
\tag{A.6}
\label{eq:chi_expand_cmp}
\]
and the same estimate yields
\[
\|\chi_t^\star\|\le D_\chi,\qquad \forall t.
\tag{A.7}
\label{eq:Dchi_bound_cmp}
\]
\paragraph{Step 2: Truncation tails for the algorithm and comparator.}
Recall the algorithm-side actual-window truncation
\[
\hat\chi_t
:=
\sum_{i=0}^{H-1}A^i(Bu_{t-1-i}+Ed_{t-1-i}).
\tag{A.9}
\label{eq:chihat_alg}
\]
By the unrolled state
representation in \eqref{eq:chi_expand_alg}, the truncation error is zero when
\(t\le H\), while for \(t>H\) it satisfies
\[
\chi_t-\hat\chi_t=A^H\chi_{t-H}.
\]
Hence, by Assumption~\ref{ass:stable_dynamics} and
\eqref{eq:Dchi_bound_alg},
\[
\|\chi_t-\hat\chi_t\|
\le
C_A\rho^H D_\chi.
\tag{A.10}
\label{eq:alg_tail_bound}
\]
Using Assumption~\ref{ass:state_lipschitz_real_cost}, we obtain
\[
|c_t(\chi_t,u_t)-c_t(\hat\chi_t,u_t)|
\le
L_\chi C_A D_\chi \rho^H.
\]
Summing over \(t\) gives
\[
J_T(u_{1:T})
\le
\sum_{t=1}^T c_t(\hat\chi_t,u_t)
+
T L_\chi C_A D_\chi \rho^H.
\tag{A.11}
\label{eq:J_alg_to_actual_window}
\]

The same argument applies to the comparator trajectory. Define
\[
\hat\chi_t^\star
:=
\sum_{i=0}^{H-1}A^i(Bu_{t-1-i}^\star+Ed_{t-1-i}).
\tag{A.12}
\label{eq:chihat_cmp}
\]
Using \eqref{eq:chi_expand_cmp} and \eqref{eq:Dchi_bound_cmp}, we similarly have
\[
\|\chi_t^\star-\hat\chi_t^\star\|
\le
C_A\rho^H D_\chi,
\]
and therefore
\[
J_T(u_{1:T}^\star)
\ge
\sum_{t=1}^T c_t(\hat\chi_t^\star,u_t^\star)
-
T L_\chi C_A D_\chi \rho^H.
\tag{A.13}
\label{eq:J_cmp_to_actual_window}
\]
\paragraph{Step 3: Memory gaps between actual-window and diagonal surrogates.}
We next compare the finite-window costs with the diagonal surrogate costs. From
the projected update and the local gradient bound in
Assumption~\ref{ass:bounded_gradients_smoothness},
\[
\|\bar u_{b+1}-\bar u_b\|
\le
\eta_B
\left\|
\frac1m
\sum_{i\in S_b}
\sum_{t\in\mathcal I_b}
\nabla c_{i,t}\bigl(\bar u_b\bigr)
\right\|
\le
\eta_B K G_{\mathrm{loc}} .
\tag{A.14}
\label{eq:block_movement_bound}
\]
Since the decision is held fixed within each block, this also implies
\[
\|u_{t+1}-u_t\|\le \eta_B K G_{\mathrm{loc}},
\qquad \forall t.
\tag{A.15}
\label{eq:pointwise_movement_bound}
\]

For the state part, the definitions of \(\hat\chi_t\) and \(\bar\chi_t\)
give
\[
\hat\chi_t-\bar\chi_t
=
\sum_{i=0}^{H-1}A^iB(u_{t-1-i}-u_t).
\]
By telescoping and \eqref{eq:pointwise_movement_bound},
\[
\|u_{t-1-i}-u_t\|
\le
\sum_{j=0}^{i}\|u_{t-j}-u_{t-j-1}\|
\le
(i+1)\eta_B K G_{\mathrm{loc}} .
\]
Using Assumption~\ref{ass:stable_dynamics}, we obtain
\[
\begin{aligned}
\|\hat\chi_t-\bar\chi_t\|
\le{}&
C_A\|B\|\eta_B K G_{\mathrm{loc}}
\sum_{i=0}^{H-1}(i+1)\rho^i\\
&\le
\frac{C_A\|B\|}{(1-\rho)^2}\eta_B K G_{\mathrm{loc}}
=
C_{\mathrm{mem}}\eta_B K G_{\mathrm{loc}} .
\end{aligned}
\tag{A.16}
\label{eq:alg_memory_gap_state}
\]
By Assumption~\ref{ass:state_lipschitz_real_cost},
\[
|c_t(\hat\chi_t,u_t)-c_t(\bar\chi_t(u_t),u_t)|
\le
L_\chi C_{\mathrm{mem}}\eta_B K G_{\mathrm{loc}} .
\]
Summing over \(t\) yields
\[
\sum_{t=1}^T c_t(\hat\chi_t,u_t)
\le
\sum_{t=1}^T c_t(\bar\chi_t(u_t),u_t)
+
T L_\chi C_{\mathrm{mem}}\eta_B K G_{\mathrm{loc}} .
\tag{A.17}
\label{eq:actual_window_to_diagonal_alg}
\]
Combining this inequality with \eqref{eq:J_alg_to_actual_window} gives
\[
\begin{aligned}
J_T(u_{1:T})
\le{}&
\sum_{t=1}^T c_t(\bar\chi_t(u_t),u_t)
+
T L_\chi C_{\mathrm{mem}}\eta_B K G_{\mathrm{loc}}\\
&+
T L_\chi C_A D_\chi \rho^H .
\end{aligned}
\tag{A.18}
\label{eq:J_alg_to_diagonal}
\]

Apply the same stable-convolution argument to the comparator. Let
\(\Delta_1^\star:=0\) and
\(\Delta_t^\star:=\|u_t^\star-u_{t-1}^\star\|\) for \(t=2,\dots,T\). Since
\(u_{1:T}^\star\in\mathcal C(V_T)\), we have
\[
\sum_{t=1}^T \Delta_t^\star \le V_T .
\tag{A.19}
\label{eq:path_length_as_deltas}
\]
Applying the preceding argument to \(u_{1:T}^\star\), with
\(\sum_{t=1}^T\Delta_t^\star\le V_T\) replacing
\eqref{eq:pointwise_movement_bound}, and using
Assumption~\ref{ass:state_lipschitz_real_cost}, gives
\[
\sum_{t=1}^T c_t(\hat\chi_t^\star,u_t^\star)
\ge
\sum_{t=1}^T c_t(\bar\chi_t(u_t^\star),u_t^\star)
-
L_\chi C_{\mathrm{mem}}V_T .
\tag{A.20}
\label{eq:cmp_actual_window_to_diagonal}
\]
Combining this with \eqref{eq:J_cmp_to_actual_window} yields
\[
\begin{aligned}
J_T(u_{1:T}^\star)
\ge{}&
\sum_{t=1}^T c_t(\bar\chi_t(u_t^\star),u_t^\star)
-
L_\chi C_{\mathrm{mem}}V_T\\
&-
T L_\chi C_A D_\chi \rho^H .
\end{aligned}
\tag{A.21}
\label{eq:J_cmp_to_diagonal}
\]
\paragraph{Step 4: Block-start comparator variation.}
To compare block-held decisions with the round-wise comparator, let
\(s_b:=(b-1)K+1\) be the first time index of block \(b\). The block-start
comparator satisfies
\[
\sum_{b=1}^{B_T-1}\|u_{s_{b+1}}^\star-u_{s_b}^\star\|
\le
\sum_{t=2}^{T}\|u_t^\star-u_{t-1}^\star\|
\le
V_T.
\tag{A.23}
\label{eq:block_cmp_variation}
\]
\paragraph{Step 5: Aligning the round-wise comparator with the block structure.}
By Assumption~\ref{ass:bounded_gradients_smoothness}, each mapping \(u\mapsto c_t(\bar\chi_t(u),u)\) is
\(G_\beta\)-Lipschitz on \(\mathcal Z\). Hence, for \(t\in\mathcal I_b\),
\[
\begin{aligned}
&c_t(\bar\chi_t(u_{s_b}^\star),u_{s_b}^\star)
-
c_t(\bar\chi_t(u_t^\star),u_t^\star)\\
&\le
G_\beta\|u_{s_b}^\star-u_t^\star\|\\
&\le
G_\beta\sum_{r=s_b+1}^{t}\|u_r^\star-u_{r-1}^\star\|.
\end{aligned}
\]
Summing this inequality over each block and then over all blocks gives
\[
\begin{aligned}
&\sum_{b=1}^{B_T}
\sum_{t\in\mathcal I_b}
\Bigl(
c_t(\bar\chi_t(u_{s_b}^\star),u_{s_b}^\star)
-
c_t(\bar\chi_t(u_t^\star),u_t^\star)
\Bigr)\\
&\le
G_\beta K V_T.
\end{aligned}
\tag{A.24}
\label{eq:block_cmp_approx}
\]
Since Algorithm~\ref{alg:block_periodic_update} holds \(u_t=\bar u_b\) on
\(\mathcal I_b\), we obtain
\[
\begin{aligned}
&\sum_{t=1}^T
\Bigl(
c_t(\bar\chi_t(u_t),u_t)
-
c_t(\bar\chi_t(u_t^\star),u_t^\star)
\Bigr)\\
&\le
\sum_{b=1}^{B_T}
\left[
c_b(\bar u_b)-c_b(u_{s_b}^\star)
\right]
+
G_\beta K V_T.
\end{aligned}
\tag{A.25}
\label{eq:round_cmp_to_block_cmp}
\]
Here \(c_b(u):=\sum_{t\in\mathcal I_b}c_t(\bar\chi_t(u),u)\).
\paragraph{Step 6: Global block smoothness.}
By Assumption~\ref{ass:bounded_gradients_smoothness} and the decomposition
\(c_t(\bar\chi_t(u),u)=N^{-1}\sum_{i=1}^N c_{i,t}(u)\), each mapping \(u\mapsto c_t(\bar\chi_t(u),u)\) is
\(L_\beta\)-smooth on \(\mathcal Z\). Therefore, the block surrogate
\(c_b\) satisfies
\[
\|\nabla c_b(u)-\nabla c_b(u')\|
\le
K L_\beta \|u-u'\|,
\qquad \forall u,u'\in\mathcal Z.
\tag{A.26}
\label{eq:block_smoothness}
\]
\paragraph{Step 7: Unbiasedness and variance of the sampled block gradient.}
For any \(u\in\mathcal Z\), the block surrogate satisfies
\[
\nabla c_b(u)
=
\frac1N\sum_{i=1}^N
\sum_{t\in\mathcal I_b}\nabla c_{i,t}(u).
\]
Therefore, since \(S_b\) is sampled uniformly without replacement, the sampled
client average is conditionally unbiased:
Here, \(\mathcal F_b\) denotes the information available after block \(b\) has
been realized and before the random subset \(S_b\) is sampled; conditional on
\(\mathcal F_b\), the local block-gradient vectors form a fixed finite
population and the only remaining randomness is the uniform sampling of
\(S_b\).
\[
\mathbb E\!\left[
\frac1m\sum_{i\in S_b}\sum_{t\in\mathcal I_b}
\nabla c_{i,t}\bigl(\bar u_b\bigr)
\middle| \mathcal F_b
\right]
=
\nabla c_b\bigl(\bar u_b\bigr).
\tag{A.27}
\label{eq:block_gradient_unbiased}
\]

It remains to bound the corresponding sampling variance. By the standard
finite-population variance identity for uniform sampling without replacement,
\[
\begin{aligned}
&\mathbb E\!\left[
\left\|
\frac1m\sum_{i\in S_b}\sum_{t\in\mathcal I_b}
\nabla c_{i,t}\bigl(\bar u_b\bigr)
-
\nabla c_b\bigl(\bar u_b\bigr)
\right\|^2
\middle| \mathcal F_b
\right] \\
&\le
\frac{N-m}{m(N-1)}
\cdot
\frac1N\sum_{i=1}^N
\left\|
\sum_{t\in\mathcal I_b}
\begin{aligned}[t]
\Bigl(&
\nabla c_{i,t}\bigl(\bar u_b\bigr)\\
&-
\frac{1}{N}\sum_{j=1}^N
\nabla c_{j,t}\bigl(\bar u_b\bigr)
\Bigr)
\end{aligned}
\right\|^2 .
\end{aligned}
\]
Using Cauchy--Schwarz over the \(K\)-round block and
Assumption~\ref{ass:gradient_heterogeneity},
\[
\frac1N\sum_{i=1}^N
\left\|
\sum_{t\in\mathcal I_b}
\begin{aligned}[t]
\Bigl(&
\nabla c_{i,t}\bigl(\bar u_b\bigr)\\
&-
\frac{1}{N}\sum_{j=1}^N
\nabla c_{j,t}\bigl(\bar u_b\bigr)
\Bigr)
\end{aligned}
\right\|^2
\le
K^2\sigma_{\mathrm{het}}^2.
\]
Consequently,
\[
\begin{aligned}
&\mathbb E\!\left[
\left\|
\frac1m\sum_{i\in S_b}\sum_{t\in\mathcal I_b}
\nabla c_{i,t}\bigl(\bar u_b\bigr)
-
\nabla c_b\bigl(\bar u_b\bigr)
\right\|^2
\middle| \mathcal F_b
\right]\\
&\le
\frac{N-m}{m(N-1)}K^2\sigma_{\mathrm{het}}^2.
\end{aligned}
\tag{A.28}
\label{eq:block_gradient_variance}
\]

\paragraph{Step 8: Block-level online analysis with changing comparators.}
We now apply the standard projected-gradient one-step inequality to the block
objective \(c_b\), with comparator \(u_{s_b}^\star\) and update
direction
\[
\frac1m\sum_{i\in S_b}\sum_{t\in\mathcal I_b}
\nabla c_{i,t}\bigl(\bar u_b\bigr).
\]
Together with convexity of \(c_b\), the conditional unbiasedness
\eqref{eq:block_gradient_unbiased}, the variance bound
\eqref{eq:block_gradient_variance}, this gives
\[
\begin{aligned}
&\mathbb E\!\left[
c_b(\bar u_b)-c_b(u_{s_b}^\star)
\,\middle|\,\mathcal F_b
\right]\\
&\le
\frac{\|\bar u_b-u_{s_b}^\star\|^2
-\mathbb E\|\bar u_{b+1}-u_{s_b}^\star\|^2}{2\eta_B} \\
&\quad+
\frac{\eta_B}{2}K^2
\left(
G_\beta^2+\frac{N-m}{m(N-1)}\sigma_{\mathrm{het}}^2
\right)
\!.
\end{aligned}
\tag{A.29}
\label{eq:one_block_master}
\]
Here the second-moment term follows from
\(\|\nabla c_b(\bar u_b)\|\le KG_\beta\) and
\eqref{eq:block_gradient_variance}.

It remains to sum the potential term over blocks. The changing comparator
satisfies
\[
\sum_{b=1}^{B_T}
\bigl(
\|\bar u_b-u_{s_b}^\star\|^2
-
\|\bar u_{b+1}-u_{s_b}^\star\|^2
\bigr)
\le
D^2+2DV_T,
\tag{A.30}
\label{eq:changing_comparator_telescope}
\]
because all points lie in \(\mathcal Z\) and
\(\sum_{b=1}^{B_T-1}\|u_{s_{b+1}}^\star-u_{s_b}^\star\|\le V_T\) by
\eqref{eq:block_cmp_variation}. Summing \eqref{eq:one_block_master} over
\(b\), using \eqref{eq:changing_comparator_telescope}, and using
\(B_TK^2\le 2KT\), we obtain
\[
\begin{aligned}
&\mathbb E\sum_{b=1}^{B_T}
\bigl[c_b(\bar u_b)-c_b(u_{s_b}^\star)\bigr]
\\
&\le
\frac{D^2}{2\eta_B}
+
\frac{D}{\eta_B}V_T \\
&\quad+
2\eta_BKT
\left[
\frac12\left(
G_\beta^2+\frac{N-m}{m(N-1)}\sigma_{\mathrm{het}}^2
\right)
\right].
\end{aligned}
\tag{A.31}
\label{eq:diagonal_regret_against_block_cmp}
\]
Finally, combining this block-level bound with the alignment inequality
\eqref{eq:round_cmp_to_block_cmp} yields
\[
\begin{aligned}
&\mathbb E\sum_{t=1}^T
\Bigl[
c_t(\bar\chi_t(u_t),u_t)
-
c_t(\bar\chi_t(u_t^\star),u_t^\star)
\Bigr]
\\
&\le
\frac{D^2}{2\eta_B}
+
\frac{D}{\eta_B}V_T \\
&\quad+
2\eta_BKT
\left[
\frac12\left(
G_\beta^2+\frac{N-m}{m(N-1)}\sigma_{\mathrm{het}}^2
\right)
\right]\\
&\quad
+
G_\beta K V_T.
\end{aligned}
\tag{A.32}
\label{eq:diagonal_regret_against_roundwise_cmp}
\]
\paragraph{Step 9.}
Combining the learner-side, comparator-side, and diagonal-regret bounds
\eqref{eq:J_alg_to_diagonal}, \eqref{eq:J_cmp_to_diagonal}, and
\eqref{eq:diagonal_regret_against_roundwise_cmp}, we obtain
\[
\begin{aligned}
&\mathbb E\!\left[
J_T(u_{1:T})-J_T(u_{1:T}^\star)
\right]
\\
&\le
\frac{D^2}{2\eta_B}
+
\frac{D}{\eta_B}V_T \\
&\quad+
2\eta_BKT
\left[
\frac12\left(
G_\beta^2+\frac{N-m}{m(N-1)}\sigma_{\mathrm{het}}^2
\right)
\right] \\
&\quad+
T L_\chi C_{\mathrm{mem}}\eta_B K G_{\mathrm{loc}}
+
(G_\beta K+L_\chi C_{\mathrm{mem}})V_T\\
&\quad+
2TL_\chi C_A D_\chi \rho^H .
\end{aligned}
\]
By the definition of \(\Lambda_{\mathrm{blk}}\), the two \(\eta_BKT\)-order
terms are bounded by \(2\eta_BKT\Lambda_{\mathrm{blk}}\). Hence
\[
\begin{aligned}
&\mathbb E\!\left[
J_T(u_{1:T})-J_T(u_{1:T}^\star)
\right]\\
&\le{}
\frac{D^2}{2\eta_B}
+
\frac{D}{\eta_B}V_T\\
&+
2\eta_BKT\Lambda_{\mathrm{blk}}
+
(G_\beta K+L_\chi C_{\mathrm{mem}})V_T\\
&+
2TL_\chi C_A D_\chi \rho^H,
\end{aligned}
\]
which proves \eqref{eq:formal_dyn_regret_pre_tuning}.

With
\(\eta_B=D/(2\sqrt{KT\Lambda_{\mathrm{blk}}})\) and
\(H=\lceil \log T/|\log\rho|\rceil\), we have
\(\rho^H\le 1/T\), and the preceding bound becomes
\[
\begin{aligned}
&\mathbb E\!\left[
J_T(u_{1:T})-J_T(u_{1:T}^\star)
\right]\\
&\le{}
2(D+V_T)\sqrt{KT\Lambda_{\mathrm{blk}}}\\
&+
(G_\beta K+L_\chi C_{\mathrm{mem}})V_T
+
2L_\chi C_A D_\chi,
\end{aligned}
\]
which proves \eqref{eq:formal_dyn_regret_post_tuning}. Since the inequality
holds for every \(u_{1:T}^\star\in\mathcal C(V_T)\), taking the minimum over
\(\mathcal C(V_T)\) gives \eqref{eq:formal_dyn_regret_final}.
\end{proof}

The main-text constants \(C_0,\dots,C_5\) in
Theorem~\ref{thm:real_dyn_regret_informal} are generic positive constants
obtained from the explicit coefficients in
\eqref{eq:formal_dyn_regret_pre_tuning}. In particular, the
optimization coefficient \(C_2\) corresponds to the coefficient multiplying
\(\eta_BKT\), namely the term involving \(\Lambda_{\mathrm{blk}}\).

\section{Proof of the Prediction-Augmented Extension}
\label{app:proof_prediction_augmented}

We first state the formal version of the prediction-augmented guarantee. Recall
\[
R=\max_{u\in\mathcal Z}\|u\|,
\qquad
D=\max_{u,u'\in\mathcal Z}\|u-u'\|.
\]
Define
\[
C_{\mathrm{mem}}
:=
\frac{C_A\|B\|}{(1-\rho)^2},
\qquad
D_\chi
:=
\frac{C_A}{1-\rho}
\bigl(\|B\|R+\|E\|D_d\bigr).
\]

\begin{theorem}[Formal bound for \textsc{OCO-S$^2$-OGD-P}]
\label{thm:prediction_augmented_formal}
Under the standing assumptions, fix a prediction sequence
\(\{M_b\}_{b=1}^{B_T}\) formed before sampling \(S_b\), and suppose that
\[
\|M_b\|\le KG_P,
\qquad b=1,\dots,B_T .
\]
Starting from \(q_1\in\mathcal Z\), define
\[
\bar u_b
:=
\Pi_{\mathcal Z}\bigl(q_b-\eta_BM_b\bigr),
\]
\[
q_{b+1}
:=
\Pi_{\mathcal Z}\left(
q_b
-
\eta_B
\frac1m
\sum_{i\in S_b}
\sum_{t\in\mathcal I_b}
\nabla c_{i,t}(\bar u_b)
\right),
\]
and set \(u_t=\bar u_b\) for all \(t\in\mathcal I_b\). Define the cumulative
prediction mismatch by
\[
\mathcal E_T(M)
:=
\sum_{b=1}^{B_T}
\mathbb E
\left\|
\frac1N
\sum_{i=1}^N
\sum_{t\in\mathcal I_b}
\nabla c_{i,t}(\bar u_b)
-
M_b
\right\|^2 .
\]
Let
\[
G_{\mathrm{ext}}
:=
G_{\mathrm{loc}}+2G_P,
\]
and
\[
\Lambda_{\mathrm{ext}}
:=
\frac12
\frac{N-m}{m(N-1)}
\sigma_{\mathrm{het}}^2
+
\frac12
L_\chi C_{\mathrm{mem}}G_{\mathrm{ext}} .
\]
Then every online decision satisfies \(u_t\in\mathcal Z\), and for every
comparator sequence \(u_{1:T}^\star\in\mathcal C(V_T)\),
\[
\begin{aligned}
&\mathbb E\!\left[
J_T(u_{1:T})-J_T(u_{1:T}^\star)
\right]\\
&\le{}
\frac{D^2}{2\eta_B}
+
\frac{D}{\eta_B}V_T\\
&+
\frac{\eta_B}{2}\mathcal E_T(M)
+
2\eta_BKT\Lambda_{\mathrm{ext}}
\\
&+
(G_\beta K+L_\chi C_{\mathrm{mem}})V_T
+
2TL_\chi C_A D_\chi \rho^H .
\end{aligned}
\label{eq:prediction-formal-untuned}
\]
In particular, if
\[
\eta_B
=
\frac{1}{\sqrt{KT}},
\qquad
H
=
\left\lceil\frac{\log T}{|\log\rho|}\right\rceil,
\]
then
\[
\begin{aligned}
&\mathbb E\!\left[
J_T(u_{1:T})-J_T(u_{1:T}^\star)
\right]\\
&\le{}
\left(\frac{D^2}{2}+2\Lambda_{\mathrm{ext}}\right)\sqrt{KT}\\
&+
D V_T\sqrt{KT}
+
\frac{\mathcal E_T(M)}{2\sqrt{KT}}\\
&+
(G_\beta K+L_\chi C_{\mathrm{mem}})V_T
+
2L_\chi C_A D_\chi .
\end{aligned}
\label{eq:prediction-formal-tuned-comparator}
\]
Consequently,
\[
\begin{aligned}
&\mathbb E\,\mathrm{Reg}_{T}(V_T)\\
&\le{}
\left(\frac{D^2}{2}+2\Lambda_{\mathrm{ext}}\right)\sqrt{KT}\\
&+
D V_T\sqrt{KT}
+
\frac{\mathcal E_T(M)}{2\sqrt{KT}}\\
&+
(G_\beta K+L_\chi C_{\mathrm{mem}})V_T
+
2L_\chi C_A D_\chi .
\end{aligned}
\label{eq:prediction-formal-tuned-regret}
\]
\end{theorem}

\begin{proof}
The feasibility claim follows directly from projection. Since
\(q_1\in\mathcal Z\), the anchor recursion gives \(q_b\in\mathcal Z\) for all
\(b\). Moreover,
\[
\bar u_b
=
\Pi_{\mathcal Z}(q_b-\eta_BM_b)
\in\mathcal Z .
\]
Thus \(u_t=\bar u_b\in\mathcal Z\) for all \(t\in\mathcal I_b\).

We next control the movement of the deployed block decisions. Since projection
is nonexpansive and \(\|M_b\|\le KG_P\),
\[
\begin{aligned}
\|\bar u_b-q_b\|
&=
\|\Pi_{\mathcal Z}(q_b-\eta_BM_b)-\Pi_{\mathcal Z}(q_b)\|
\\
&\le
\eta_BKG_P .
\end{aligned}
\]
The anchor update satisfies
\[
\begin{aligned}
\|q_{b+1}-q_b\|
&\le
\eta_B
\left\|
\frac1m
\sum_{i\in S_b}
\sum_{t\in\mathcal I_b}
\nabla c_{i,t}(\bar u_b)
\right\|  \\
&\le
\eta_B K G_{\mathrm{loc}},
\end{aligned}
\]
where the last inequality uses the bounded-gradient assumption. Hence
\[
\begin{aligned}
\|\bar u_{b+1}-\bar u_b\|
&\le
\|\bar u_{b+1}-q_{b+1}\|
+
\|q_{b+1}-q_b\|\\
&\quad+
\|q_b-\bar u_b\| \\
&\le
\eta_BK(G_{\mathrm{loc}}+2G_P)
=
\eta_BK G_{\mathrm{ext}} .
\end{aligned}
\]
Therefore the algorithm-side memory-gap argument used for
\textsc{OCO-S$^2$-OGD} applies with \(G_{\mathrm{loc}}\) replaced by
\(G_{\mathrm{ext}}\), yielding
\[
\begin{aligned}
J_T(u_{1:T})
\le{}&
\sum_{t=1}^T c_t(\bar\chi_t(u_t),u_t)
+
TL_\chi C_{\mathrm{mem}}\eta_BK G_{\mathrm{ext}}\\
&+
TL_\chi C_A D_\chi\rho^H .
\end{aligned}
\label{eq:prediction-alg-side-reduction}
\]

It remains to control the diagonal surrogate regret. Let
\(s_b:=\min\mathcal I_b\), and define the sampled block gradient
\[
\widehat G_b
:=
\frac1m
\sum_{i\in S_b}
\sum_{t\in\mathcal I_b}
\nabla c_{i,t}(\bar u_b),
\]
and the full block-level surrogate gradient
\[
G_b
:=
\frac1N
\sum_{i=1}^N
\sum_{t\in\mathcal I_b}
\nabla c_{i,t}(\bar u_b).
\]
By the update rule,
\[
q_{b+1}
=
\Pi_{\mathcal Z}(q_b-\eta_B\widehat G_b),
\qquad
\bar u_b
=
\Pi_{\mathcal Z}(q_b-\eta_BM_b).
\]
The standard optimistic projected-gradient inequality gives, for any
\(v\in\mathcal Z\),
\[
\begin{aligned}
&\langle \widehat G_b,\bar u_b-v\rangle\\
&\le{}
\frac{\|q_b-v\|^2-\|q_{b+1}-v\|^2}{2\eta_B}
+
\frac{\eta_B}{2}\|\widehat G_b-M_b\|^2\\
&-
\frac{\eta_B}{2}\|\widehat G_b-G_b\|^2
\\
&+
\frac{\eta_B}{2}\|G_b-M_b\|^2 .
\end{aligned}
\]
Taking \(v=u_{s_b}^\star\), using convexity of each diagonal surrogate
component, and taking conditional expectation with respect to the randomness of
\(S_b\), we obtain
\[
\begin{aligned}
&\mathbb E\!\left[
\sum_{t\in\mathcal I_b}
c_t(\bar\chi_t(\bar u_b),\bar u_b)
-
\sum_{t\in\mathcal I_b}
c_t(\bar\chi_t(u_{s_b}^\star),u_{s_b}^\star)
\,\middle|\,\mathcal F_b
\right] \\
&\le
\frac{
\|q_b-u_{s_b}^\star\|^2
-
\mathbb E\|q_{b+1}-u_{s_b}^\star\|^2
}{2\eta_B}
+
\frac{\eta_B}{2}
\|G_b-M_b\|^2  \\
&\quad+
\frac{\eta_B}{2}
K^2
\frac{N-m}{m(N-1)}
\sigma_{\mathrm{het}}^2 .
\end{aligned}
\label{eq:prediction-block-descent}
\]
Here we used the conditional unbiasedness of \(\widehat G_b\) and the
finite-population sampling-variance bound
\[
\mathbb E\!\left[
\|\widehat G_b-G_b\|^2
\,\middle|\,\mathcal F_b
\right]
\le
K^2
\frac{N-m}{m(N-1)}
\sigma_{\mathrm{het}}^2 .
\]

Summing \eqref{eq:prediction-block-descent} over blocks and applying the usual
changing-comparator telescope gives
\[
\begin{aligned}
&\mathbb E
\sum_{b=1}^{B_T}
\left[
\sum_{t\in\mathcal I_b}
c_t(\bar\chi_t(\bar u_b),\bar u_b)
-
\sum_{t\in\mathcal I_b}
c_t(\bar\chi_t(u_{s_b}^\star),u_{s_b}^\star)
\right] \\
&\le
\frac{D^2}{2\eta_B}
+
\frac{D}{\eta_B}V_T\\
&\quad+
\frac{\eta_B}{2}\mathcal E_T(M)
+
2\eta_BKT
\left[
\frac12
\frac{N-m}{m(N-1)}
\sigma_{\mathrm{het}}^2
\right],
\end{aligned}
\label{eq:prediction-block-sum}
\]
where we used \(B_TK^2\le 2KT\).

Next, the same block-alignment argument as in the baseline analysis gives
\[
\begin{aligned}
&\mathbb E
\sum_{t=1}^T
\left[
c_t(\bar\chi_t(u_t),u_t)
-
c_t(\bar\chi_t(u_t^\star),u_t^\star)
\right] \\
&\le
\frac{D^2}{2\eta_B}
+
\frac{D}{\eta_B}V_T\\
&\quad+
\frac{\eta_B}{2}\mathcal E_T(M)
+
2\eta_BKT
\left[
\frac12
\frac{N-m}{m(N-1)}
\sigma_{\mathrm{het}}^2
\right]\\
&\quad
+
G_\beta K V_T .
\end{aligned}
\label{eq:prediction-diagonal-regret}
\]

Finally, combine the algorithm-side diagonal reduction
\eqref{eq:prediction-alg-side-reduction}, the comparator-side diagonal
reduction, and \eqref{eq:prediction-diagonal-regret}. The comparator-side
reduction contributes
\[
L_\chi C_{\mathrm{mem}}V_T
+
TL_\chi C_A D_\chi\rho^H .
\]
Therefore,
\[
\begin{aligned}
&\mathbb E\!\left[
J_T(u_{1:T})-J_T(u_{1:T}^\star)
\right]\\
&\le{}
\frac{D^2}{2\eta_B}
+
\frac{D}{\eta_B}V_T\\
&+
\frac{\eta_B}{2}\mathcal E_T(M)
+
G_\beta K V_T
+
L_\chi C_{\mathrm{mem}}V_T\\
&+
2TL_\chi C_A D_\chi\rho^H\\
&+
2\eta_BKT
\left[
\frac12
\frac{N-m}{m(N-1)}
\sigma_{\mathrm{het}}^2
+
\frac12
L_\chi C_{\mathrm{mem}}G_{\mathrm{ext}}
\right].
\end{aligned}
\]
By the definition of \(\Lambda_{\mathrm{ext}}\), this is exactly
\eqref{eq:prediction-formal-untuned}.

For the fixed-stepsize bound, substitute
\[
\eta_B
=
\frac{1}{\sqrt{KT}}
\]
into the first four terms of \eqref{eq:prediction-formal-untuned}. This gives
\[
\begin{aligned}
&\frac{D^2}{2\eta_B}
+
\frac{D}{\eta_B}V_T
\\
&\quad
+
\frac{\eta_B}{2}\mathcal E_T(M)
+
2\eta_BKT\Lambda_{\mathrm{ext}}
\\
&=
\left(\frac{D^2}{2}+2\Lambda_{\mathrm{ext}}\right)\sqrt{KT}
\\
&\quad+
D V_T\sqrt{KT}
+
\frac{\mathcal E_T(M)}{2\sqrt{KT}}.
\end{aligned}
\]
Moreover, with
\[
H
=
\left\lceil\frac{\log T}{|\log\rho|}\right\rceil,
\]
we have \(T\rho^H\le 1\), and hence
\[
2TL_\chi C_A D_\chi\rho^H
\le
2L_\chi C_A D_\chi .
\]
This proves \eqref{eq:prediction-formal-tuned-comparator}. Since the bound
holds for every \(u_{1:T}^\star\in\mathcal C(V_T)\), taking the minimum over
\(\mathcal C(V_T)\) gives \eqref{eq:prediction-formal-tuned-regret}.
\end{proof}

The constants \(C_0,C_1,C_2^{\mathrm{res}},C_3,C_4,C_5,C_6\) used in
Theorem~\ref{thm:prediction_augmented_regret} are generic positive constants
obtained from the explicit coefficients in
\eqref{eq:prediction-formal-untuned}. In particular,
\(C_2^{\mathrm{res}}\) corresponds to the residual coefficient multiplying
\(\eta_BKT\), while \(C_6\) corresponds to the coefficient multiplying
\(\eta_B\mathcal E_T(M)\).